\documentclass[aps,prl,twocolumn,showpacs]{revtex4}
\usepackage{graphicx,amsfonts,amsmath,amssymb,epsfig}
\begin{document}

\title{Photoassociation spectroscopy of a Spin-1 Bose-Einstein condensate}
\author{C.D. Hamley, E.M. Bookjans, G. Behin-Aein, P. Ahmadi, and M.S. Chapman}

\affiliation{School of Physics, Georgia Institute of Technology,
  Atlanta, GA 30332-0430 }

\begin{abstract}
We report on the high resolution photoassociation spectroscopy of a
$^{87}$Rb spin-1 Bose-Einstein condensate to the $1_\mathrm{g}
(P_{3/2})~ v = 152$ excited molecular states. We demonstrate the use
of spin dependent photoassociation to experimentally identify the
molecular states and their corresponding initial scattering channel.
These identifications are in excellent agreement with the
eigenvalues of a hyperfine-rotational Hamiltonian. Using the
observed spectra we estimate the change in scattering length and
identify photoassociation laser light frequency ranges that maximize
the change in the spin-dependent mean-field interaction energy.
\end{abstract}

\maketitle

In a spinor Bose-Einstein condensate \cite{1,2,3,4}, the delicate
interplay between different atomic spin orientations results in a
small spin dependence of the collisional interaction energy
\cite{5}. Though the spin dependence is small relative to the total
interaction energy ({\it e.g.}\thinspace $\sim$ 0.5$\%$ in
$^{87}$Rb), the coherence of the condensate allows observation of
novel phenomena such as coherent spin mixing and dynamics
\cite{6,7,8,9,10}, metastable states \cite{11}, spin domains
\cite{12, 13} and quantum tunneling across spin domains \cite{14}.
The spin-dependent interaction energy arises from the small
difference in the $s$-wave scattering lengths of the allowed angular
momentum channels and manifests in anti- or ferromagnetic properties
for a spin-1 condensate depending on the algebraic sign of the
difference. In the two spin-1 condensates investigated to date,
$^{23}$Na and $^{87}$Rb, the former is anti-ferromagnetic and the
latter is ferromagnetic.

Magnetic \cite{15} and optical \cite{16} Feshbach resonances have
been employed to dynamically change the scattering lengths which
offers the tantalizing prospect of influencing the spinor dynamics
or even possibly changing the magnetic nature of the condensate.
However, spinor dynamics are suppressed for magnetic fields greater
than a few 100s of mG \cite{7, 16b}. The fields used in magnetic
Feshbach resonance experiments are much larger than this and thus
can not be used for altering the spinor dynamics. Photoassociation
around single resonance lines also has limited flexibility since the
atom loss rate is very high near the resonances where the change in
scattering lengths is appreciable. To circumvent this problem it has
been suggested to use the effect of multiple photoassociation lines
from a vibrational level with a rich hyperfine structure \cite{17}.
This allows parameters that depend on multiple scattering lengths to
be enhanced between lines from different scattering channels while
the atom loss rate is reduced. Specifically, variation of the spin
dependent interaction strength could be optimized if there are
adjacent molecular states that occur through different scattering
channels. In order to identify the existence of such lines,
collision channel selective spectroscopy needs to be performed which
requires control over the Zeeman states of the colliding atoms. Note
that although the effects of hyperfine levels of the colliding atoms
on the photoassociation spectrum has been previously observed
\cite{18, 19}, there has not been any collision channel selective
spectroscopy.

In this Letter, we report on experimental photoassociation
spectroscopy of a spinor condensate. By observing the effect of the
Zeeman states of the colliding atoms on the photoassociation
spectrum, we extract information about the angular momentum of the
molecular states.  We identify for the first time molecular states
that are allowed for a specific scattering channel while forbidden
for another.  From the spectrum, we calculate the atom loss rate and
the expected change in the mean-field interaction energy. Using the
observed spectra and these calculations, we identify suitable
photoassociation light frequencies that maximize change of the spin
dependent interaction strength with modest atom loss rates.

The experiment is performed on $^{87}$Rb condensates created
directly in an optical trap \cite{1}. After the condensate is
formed, it is exposed to a photoassociation (PA) laser. To measure
the spectrum of the molecular excited states, the condensate
population losses are measured for different frequencies of the PA
light. The PA excitation laser frequency is tuned near the
$1_\mathrm{g}(P_{3/2})~ v = 152$ state which has a rich hyperfine
structure and a binding energy of 24.1 cm$^{-1}$ \cite{20, 21} below
the D$_2$ line of $^{87}$Rb. The PA laser has a focused waist of 80
$\mu$m, and its frequency is actively stabilized with an accuracy
$\sim$5 MHz using a transfer cavity locked to a stabilized diode
laser.
\begin{figure*}
\begin{center}
\includegraphics[width=7in]{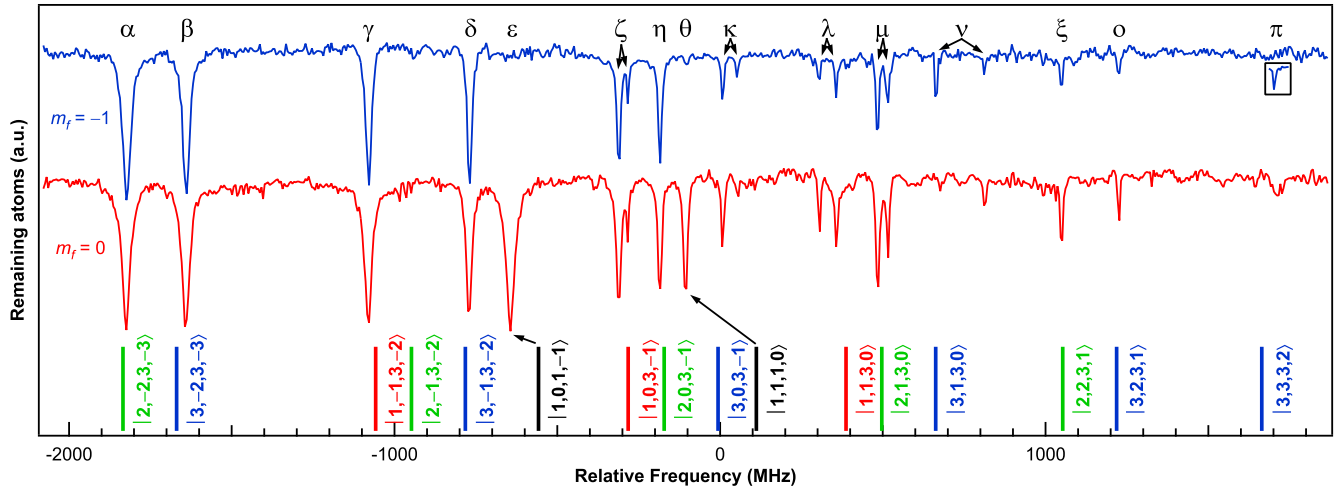}
\caption{Observed photoassociation spectrum of the $1_\mathrm{g}
(P_{3/2})~ v = 152$ state for $m_f  = -1$ (upper) and $m_f = 0$
(lower).  These spectra are obtained after 150 ms of exposing a BEC
to a PA light with 3.8 mW and 80 micron beam waist. The inset box
for $m_f  = -1$ is obtained using 350 ms and 11 mW with averaging to
enhance visibility of this weaker line. Plots are offset for
clarity. The origin of the hyperfine-rotation Hamiltonian fit is
used as the zero point for the plot. A stick spectrum with
approximate $|F,f,I,i\rangle$ labels is given.} \label{fig1}
\end{center}
\end{figure*}
Fig.\thinspace \ref{fig1} shows the observed photoassociation
spectrum of the $1_\mathrm{g}(P_{3/2})~ v = 152$ state taken using
condensates with $|f=1,m_f=-1\rangle$ and $|f=1,m_f=0\rangle$ spin
states.  A portion of the $|f=1,m_f=-1\rangle$ spectrum was observed
in \cite{21}. The PA light is ramped up to 3.8 mW in 50 ms, after
which it remains on for 100 ms. Successive data points are separated
by 5 MHz. The weaker lines identified by $\kappa $ through $\pi$
have been confirmed using more intensity and longer time but are
shown here under the same conditions as the stronger lines for
consistency. The inset shows a line too weak to be made out in the
larger scan. In order to enhance the visibility of this line, the PA
light power is increased to 11 mW and left on for 300 ms, and it has
been averaged over four scans.  For each data point, an absorptive
image of the condensate is taken 12 ms after the PA and trapping
lasers are turned off. The condensate population is counted and
normalized to an average value taken under the same conditions but
with the PA light frequency detuned far from any of the molecular
lines observed in Fig.\thinspace \ref{fig1}.

The $1_\mathrm{g}$ potential is categorized as a Hund's case (c).
The relevant good quantum numbers for this potential are total
molecular angular momentum, $F$, total nuclear spin, $I$, and
$\Omega$, the projection of the rotational angular momentum, $J$, on
the internuclear axis. For this case, the projections $F_{\bar{z}}$,
$I_{\bar{z}}$ of $\vec{F}$and $\vec{I}$ on the internuclear axis are
almost good quantum numbers, while $J$ is not \cite{22}.

The possible values of $F$ are determined by bosonic symmetry and
angular momentum addition rules. For two identical spin-1 bosons,
the allowed $s$-wave collision channels are $F = 2$ or 0. Addition
of the angular momentum of the colliding atoms and the PA photon
specifies the possible total angular momentum, $F$, of the available
molecular states for each scattering channel. Photoassociation
through the total spin 2 scattering channel gives molecular $F$
numbers of 1, 2, and 3 while the total spin 0 scattering channel
restricts $F$ to 1.

\begin{figure}
\begin{center}
\includegraphics[width=3.3in]{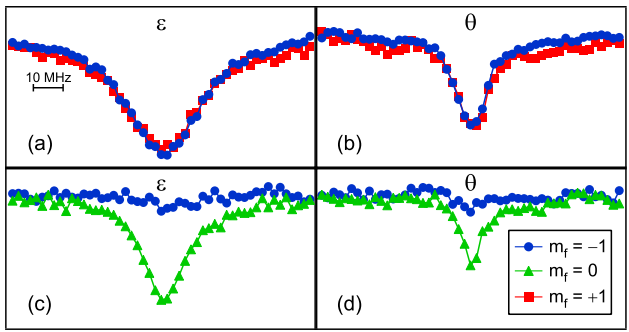}
\caption{Photoassociation spectroscopy of $\varepsilon$ and $\theta$
with different mixtures of spin states. (a) and (b) Lines
$\varepsilon$ and $\theta$ for mixture of $m_f = -1$ and $m_f = 1$
spin states. (c) and (d) for mixture of  $m_f  = -1$ and $m_f = 0$.
The observed data points corresponding to $m_f = (-1, 0, 1)$ spin
states are represented by circles, triangles, and squares
respectively.}\label{fig2}
\end{center}
\end{figure}

In the following, we use the experimental properties of the
photoassociation spectrum to identify these quantum numbers for the
observed molecular lines. The most striking difference between the
two PA spectra in Fig.\thinspace \ref{fig1} is that the lines
$\varepsilon$ and $\theta$ appear only for the condensate containing
the $m_f = 0$ spin state. Since two $m_f = -1$ atoms can only
scatter through the total spin 2 channel, while two $m_f = 0$ atoms
access both the total spin 2 and 0 channels, this observation
indicates that these lines occur through the total spin 0 scattering
channel, restricting $F$ for the molecular state represented by
these lines to 1. This conclusion is further confirmed by repeating
the same experiment with condensates containing both $m_f = -1$  and
$m_f = 1$ spin states and also with condensates containing $m_f =
-1$ and $m_f = 0$ (see Fig.\thinspace \ref{fig2}). These spin state
combinations are prepared using a microwave manipulation technique
similar to the one used in Ref. \cite{7}. The populations of
different $m_f$ states are counted by spatially separating them
using a Stern-Gerlach field during the time of flight. An $m_f = -1
(m_f = 1)$ can only access the total spin 0 scattering channel
through collision with an $m_f = 1 (m_f = -1)$ spin state.
Therefore, neither of these spin states participate in the total
spin 0 channel photoassociation if they coexist in the condensate
only with an $m_f = 0$ spin state.
To illustrate these predictions, PA spectra across the lines
$\varepsilon$ and $\theta$ are taken with condensates containing
different mixtures of spin states as shown in Fig.\thinspace
\ref{fig2}.

The data shown in Figs.\thinspace \ref{fig2}(a) and \ref{fig2}(b)
are taken with similar conditions to Fig.\thinspace \ref{fig1},
whereas in Fig.\thinspace \ref{fig2}(c) and \ref{fig2}(d) the
photoassociation power is decreased to 1.5 mW to reduce the
mechanical effects on the $m_f = -1$ atoms caused by the rapid loss
of the $m_f = 0$  atoms from the trap. These spectra are taken with
frequency steps of 2 MHz. The observed data in this figure shows
that the $m_f = -1$ and $m_f = 1$ spin states participate in the
photoassociation for the lines $\varepsilon$ and $\theta$ if both of
them coexist in the condensate, whereas if $m_f = -1$ spin state
coexists with the $m_f = 0$ spin state it does not participate in
the photoassociation process. This observation matches our analysis
based on scattering channels and confirms that $F = 1$ for lines
$\varepsilon$ and $\theta$.

\begin{figure}
\begin{center}
\includegraphics[width=3.3in]{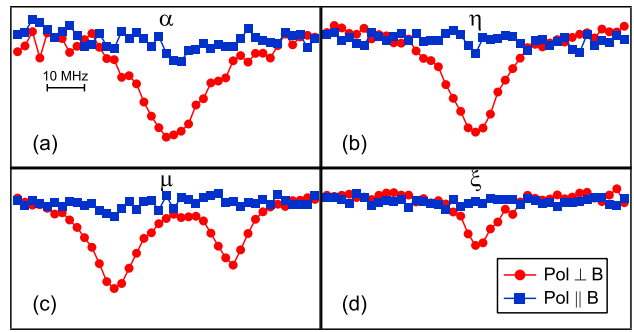}
\caption{Photoassociation spectroscopy through lines $\alpha$, $\eta
$, $\mu$, and $\xi$ for an $m_f = 0$ spin state condensate to an $F
= 2$ molecular state. The data points are taken with the
polarization vector of the PA light along and perpendicular to an
external magnetic field (squares) and (circles).} \label{fig3}
\end{center}
\end{figure}

To identify $F = 2$ PA lines we note that a $\Delta F = 0$, $\Delta
M_F = 0$ transition from a total two atom collisional spin state
$|2,0\rangle$ to a bound molecular state $|2,0\rangle$ is forbidden
by the electric dipole selection rule. Therefore the lines with $F =
2$ are suppressed for an $m_f = 0$ spin state condensate with
$\pi$-polarized PA light. Such lines are found by taking a PA
spectrum of a pure $m_f = 0$ condensate in the presence of an
external B-field of 1 G along the polarization of the PA light. The
observed data for this case, shown in Fig.\thinspace \ref{fig3},
indicates that the lines $\alpha$, $\eta $, $\mu$, and $\xi$ are
absent. By changing the polarization of the PA light perpendicular
to the quantization axis defined by the applied B-field, all of
these lines reappear in the spectrum. A comparison between the
observed data with polarization aligned parallel (square) and
perpendicular (circle) to the external B-field is shown in
Fig.\thinspace \ref{fig3}, which confirms $F = 2$ as a good quantum
number for these lines. Fig.\thinspace \ref{fig3}(c) also indicates
that the line $\mu$ is split into two $F = 2$ components.

From these observations and the spacing of the lines we are able to
assign $F$ and $I$ for the observed lines. This is accomplished by
noting that for possible values of $F =$ 1, 2, and 3 the eigenvalues
of $\vec{F}^2$ are 2, 6, and 12. Therefore the separation between
molecular states with $F = 1$ and 2 should be $2/3$ of the frequency
separation between the lines with $F = 2$ and 3. $F$ numbers for the
remaining lines are readily deduced by their spacing and order in
each hyperfine grouping. The total nuclear spin for these lines is
found by decomposing the initial scattering wave function of the two
colliding atoms. This analysis shows that the total nuclear spin
must be 1 for the total spin 0 scattering channel and predominantly
$I = 3$ for the total spin 2 scattering channel. Table\thinspace
\ref{tab:table1} gives the assigned values of $F$ and $I$ for the
observed lines.
\begin{table}
\caption{\label{tab:table1} Assigned total angular momentum and
nuclear spin of the molecular states appearing in the PA spectra of
Fig.\thinspace \ref{fig1}.}
\begin{ruledtabular}
\begin{tabular}{c|ccccccccccccccc}
 & $\alpha$&$\beta$&$\gamma$&$\delta$&$\varepsilon$&$\zeta$&$\eta$&$\theta$&$\kappa$&
 $\lambda$&$\mu$&$\nu$&$\xi$&$o$&$\pi$ \\
 \hline $F$&2&3&1&3&1&1&2&1&3&1&2&3&2&3&3\\
 \hline $I$&3&3&3&3&1&3&3&1&3&3&3&3&3&3&3
\end{tabular}
\end{ruledtabular}
\end{table}

For the $1_\mathrm{g}$ potential, the molecular hyperfine constant
is comparable to the small rotational constant of the state.
Therefore, the effective Hamiltonian for the $1_\mathrm{g}$
rotation-hyperfine structure can be written as \cite{22},
\begin{eqnarray}
H_{H-R}\!\!&=&\!\!a_v \vec{I}\!\!\cdot \!
\vec{\Omega} + B_v\vec{J}^2 \nonumber\\
&=&a_v ~ I_{\bar{z}}~ \Omega \nonumber \\
&+&\!B_v \{\vec{F}^2\!+\!\vec{I}^2\!-\!2 F_{\bar{z}}~ I_{\bar{z}} -
\bar{F}_{+}\! \bar{I}_{-}\!-\!\bar{F}_{-}\!\bar{I}_{+}\!\},
\label{Eq1}
\end{eqnarray}
where the operators are defined with respect to the inter-nuclear
axis $\bar{z}$. Since changing the signs of $F_{\bar{z}}$ ,
$I_{\bar{z}}$, and $\Omega$ results in a degenerate state, the
states are labeled by $F$, $f$, $I$, and $i$, where $f=F_{\bar{z}}
\cdot \Omega$ and $i=I_{\bar{z}} \cdot \Omega$. The eigenvalues of
the Hamiltonian are obtained by diagonalizing its corresponding
matrix in the $|F, F_{\bar{z}}, I, I_{\bar{z}}\rangle$ basis set.
The coupling is sufficiently weak that labeling the final states by
the almost good quantum numbers of $f$ and $i$ is justified. These
eigenvalues are fit to the observed spectrum of the $m_f = -1$
condensate of Fig.\thinspace \ref{fig1} using the least squares
approach with $a_v$, $B_v$, and the frequency offset
$\omega_{\mathrm{offset}}$ as fitting parameters. In case of the
split lines, the stronger one is used for fitting. Using the $a_{F =
2}$ lines the fitting specifies $a_v$ and $B_v$ to be 667(5) and
27.6(1.3) MHz, respectively. The positions of the eigenvalues using
these parameters are shown as a stick spectrum in Fig.\thinspace
\ref{fig1} which is in good agreement with the observed locations of
the atom loss lines in the PA spectrum. The stick spectrum is
labeled by the dominant part of the corresponding eigenvectors which
is in complete agreement with the assigned quantum numbers in
Table\thinspace \ref{tab:table1}.

A few general comments should be made for the observed as well as
predicted spectra. First, the molecular state corresponding to the
$|2,-1,3,-2\rangle$ eigenstate is not observed under any
experimental condition. This line is also absent in a scan of the $v
= 153$ vibrational level. Second, predicted locations of the lines
$\varepsilon$ and $\theta$ corresponding to the $a_{F = 0}$
scattering channel are in poorer agreement with their observed
locations compared to the other lines. These two lines are connected
by arrows to their assumed eigenvalues in Fig.\thinspace \ref{fig1}.
There appears to be a discrepancy both in the separation of the
lines and their apparent origin for the hyperfine-rotation
Hamiltonian. It was unclear as to whether different fit parameters
or a revision to the Hamiltonian is needed to correct the
deviations. Last, we note that some of the lines appear to be split
on the order of 30 to 160 MHz. These split lines are connected to
their labels in Fig.\thinspace \ref{fig1} for clarity. The
Hamiltonian presented in Eq.\thinspace (\ref{Eq1}) predicts that all
lines should be doubly degenerate and does not account for this
splitting. This points to an additional interaction not considered
in this simple model. These discrepancies are important for
refinement of the molecular potential theory.
\begin{figure}
\begin{center}
\includegraphics[width=3.3in]{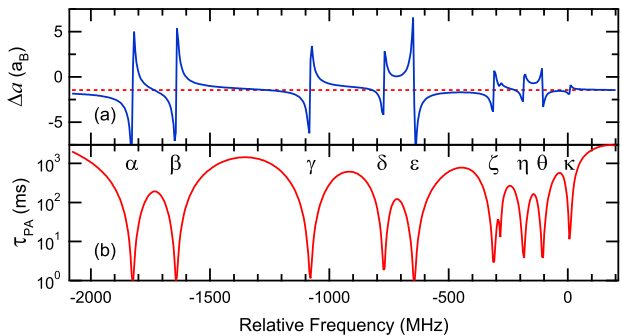}
\caption{Calculated values for $\Delta a$ and for the PA limited
lifetime of BEC. The dotted line is the nominal value of $\Delta a$.
The intensity here is 20 times that used to take the spectra in
Fig.\thinspace \ref{fig1}.} \label{fig4}
\end{center}
\end{figure}

One of our motivations for studying this system is to assess the
suitability of photoassociation induced change in two $s$-wave
scattering lengths, $\Delta a = a_{F=2} - a_{F=0}$, in order to
manipulate spin-dependent properties of a condensate. To determine
the variation of  $\Delta a$, the observed data of Fig.\thinspace
\ref{fig1} is fitted to a theoretical formula for the inelastic loss
rate ($K_{\mathrm{inel}}$) to find the width and amplitude for each
line as discussed in Ref.\thinspace \cite{16}. These fit parameters
along with the condensate density ($n_0$) are then used to calculate
the estimated change in  $\Delta a$ as shown in Fig.\thinspace
\ref{fig4}(a).

We also calculate a photoassociation limited lifetime,
$\tau_{PA}=(K_{\mathrm{inel}} \times n_0)^{-1} $, during which spin
dependent dynamics can be observed before the condensate is depleted
by the PA atom loss as shown in Fig.\thinspace \ref{fig4}(b). The
plots in Fig.\thinspace \ref{fig4} assume a PA intensity 20 times
higher than that used to measure the spectrum. One of the more
interesting frequency ranges in Fig.\thinspace \ref{fig4} is between
lines $\delta$ and $\varepsilon$ which are associated with different
scattering channels. In this frequency range the influences of the
two lines reinforce each other for a large change in $\Delta a$, but
with only a modest reduction in the 3 s lifetime of the condensate
due to the PA light. From this data it is clear that the value of
$\Delta a$ can be altered significantly from its nominal value of
1.45(32)$\mathrm{a}_B$ \cite{7} up to a total cancelation, with
sufficient PA limited lifetimes needed to see the change in spinor
dynamics. However to change the sign and hence the magnetic nature
of the condensate, the values needed correspond to challengingly
short life times compared to the time needed to relax to the
magnetic ground state. Hopefully, other molecular states will have a
slightly more favorable width, amplitude, or spacing.

In conclusion, we performed photoassociation spectroscopy on a
spin-1 condensate. The spin dependent photoassociation spectrum is
used to identify good quantum numbers for some of the molecular
states, which are in agreement with theoretical predictions of a
hyperfine-rotation Hamiltonian. The spectrum is also used to predict
the photoassociation limited lifetime and change in $\Delta a$. It
is shown that optical Feshbach resonances to a molecular state with
hyperfine structure are a viable method of altering the
spin-dependent mean-field interaction energy to change the spinor
dynamics of the condensate. The results of this study provide a
valuable test case to answer general questions about how to model
molecular potentials in the presence of hyperfine and rotation
interactions. A high resolution photoassociation study of other
vibrational levels will provide a deeper understanding of molecular
potentials. This work is supported by the National Science
Foundation, PHYS-0605049.


\begin{references}
\bibitem{1} M.D.\thinspace Barrett {\it et al.}, Phys.\thinspace Rev.\thinspace
Lett. {\bf 87}, 010404 (2001).

\bibitem{2} E.A.\thinspace Cornell {\it et al.}, J.\thinspace Low Temp.\thinspace Phys. {\bf 113}, 151 (1998).

\bibitem{3} C.J.\thinspace Myatt {\it et al.}, Phys.\thinspace Rev.\thinspace
Lett. {\bf 78}, 586 (1997).

\bibitem{4} D.M.\thinspace Stamper-Kurn {\it et al.}, Phys.\thinspace Rev.\thinspace
Lett. {\bf 80}, 2027 (1998).

\bibitem{5} T.L.\thinspace Ho, Phys.\thinspace Rev.\thinspace
Lett. {\bf 81}, 742 (1998).

\bibitem{6} M.S.\thinspace Chang {\it et al.}, Phys.\thinspace Rev.\thinspace
Lett. {\bf 92}, 140403 (2004).

\bibitem{7} M.S.\thinspace Chang {\it et al.}, Nat.\thinspace Phys. {\bf 1}, 111 (2005).

\bibitem{8} W.X.\thinspace Zhang {\it et al.}, Phys.\thinspace Rev.\thinspace A {\bf 72}, 013602 (2005).

\bibitem{9} H.\thinspace Schmaljohann {\it et al.}, Phys.\thinspace Rev.\thinspace
Lett. {\bf 92}, 040402 (2004).

\bibitem{10} A.\thinspace Widera {\it et al.}, Phys.\thinspace Rev.\thinspace
Lett. {\bf 95}, 190405 (2005).

\bibitem{11} H.J.\thinspace Miesner {\it et al.}, Phys.\thinspace Rev.\thinspace
Lett. {\bf 82}, 2228 (1999).

\bibitem{12} J.\thinspace Stenger {\it et al.}, Nature {\bf 396}, 345 (1998).

\bibitem{13} L.E.\thinspace Sadler {\it et al.}, Nature {\bf 443}, 312 (2006).

\bibitem{14} D.M.\thinspace Stamper-Kurn {\it et al.}, Phys.\thinspace Rev.\thinspace
Lett. {\bf 83}, 661 (1999).

\bibitem{15} S.\thinspace Inouye {\it et al.}, Nature {\bf 392}, 151 (1998).

\bibitem{16} M.\thinspace Theis {\it et al.}, Phys.\thinspace Rev.\thinspace
Lett. {\bf 93}, 123001 (2004).

\bibitem{16b} M.\thinspace Erhard, Appl.\thinspace Phys.\thinspace B {\bf79}, 1001
(2004).

\bibitem{17} M.W.\thinspace Jack, and M.\thinspace Yamashita,
Phys.\thinspace Rev.\thinspace A {\bf 71}, 033619 (2005).

\bibitem{18} E.R.I.\thinspace Abraham {\it et al.}, Phys.\thinspace Rev.\thinspace A {\bf 53}, 3092 (1996).

\bibitem{19}  E.\thinspace Tiesinga {\it et al.}, Phys.\thinspace Rev.\thinspace A {\bf 71}, 052703 (2005).

\bibitem{20} J.D\thinspace. Miller {\it et al.}, Phys.\thinspace Rev.\thinspace
Lett. {\bf 71}, 2204 (1993).

\bibitem{21} M.\thinspace Theis, "Optical Feshbach Resonances in a Bose-Einstein Condensate," doctoral thesis
(University of Innsbruck, 2005).

\bibitem{22}  X.T.\thinspace Wang {\it et al.}, Phys.\thinspace Rev.\thinspace A {\bf 57}, 4600 (1998).
\end{references}
\end{document}